# IDPS: An Integrated Intrusion Handling Model for Cloud Computing Environment

Hassen Mohammed Alsafi , Wafaa Mustafa Abduallah and Al-Sakib Khan Pathan

Department of Computer Science

Faculty of Information and Communication Technology

International Islamic University Malaysia (IIUM), Malaysia

abosafi87@gmail.com, heevy9@yahoo.com, and sakib@iium.edu.my

## Abstract

Today, many organizations are moving their computing services towards the Cloud. This makes their computer processing available much more conveniently to users. However, it also brings new security threats and challenges about safety and reliability. In fact, Cloud Computing is an attractive and cost-saving service for buyers as it provides accessibility and reliability options for users and scalable sales for providers. In spite of being attractive, Cloud feature poses various new security threats and challenges when it comes to deploying Intrusion Detection System (IDS) in Cloud environments. Most Intrusion Detection Systems (IDSs) are designed to handle specific types of attacks. It is evident that no single technique can guarantee protection against future attacks. Hence, there is a need for an integrated scheme which can provide robust protection against a complete spectrum of threats. On the other hand, there is great need for technology that enables the network and its hosts to defend themselves with some level of intelligence in order to accurately identify and block malicious traffic and activities. In this case, it is called Intrusion prevention system (IPS). Therefore, in this paper, we emphasize on recent implementations of IDS on Cloud Computing environments in terms of security and privacy. We propose an effective and efficient model termed as the Integrated Intrusion Detection and Prevention System (IDPS) which combines both IDS and IPS in a single mechanism. Our mechanism also integrates two techniques namely, Anomaly Detection (AD) and Signature Detection (SD) that can work in cooperation to detect various numbers of attacks and stop them through the capability of IPS.

# 1. Introduction

In the last decade, most people were concerned about obtaining computers in their offices, schools and homes. The main reason behind that was to get close to the world and communicate and exchange data via these devices. In contrast, today people are concerned about the Internet and its speed for effective and efficient communication. In addition, often they need extra services to the existing legacy service provided by the Internet. These services are known as some kinds of computing tasks that are delivered by the Internet Service Providers (ISP).

While getting required service is the users' demand, with the advanced development of the Internet tools around the world, attackers also aim to identify various loopholes in the operating system and networks. When we talk about Clouds, the main target of the attackers is to make illegitimate and unlawful attack to the available resources in the Cloud computing settings. In order to overcome these obstacles, some actions need be taken in the host based (HB) and network based (NB) level. Even though the use of intrusion detection system (IDS) is not guaranteed and cannot be considered as complete defense, we believe it can play a significant role in the Cloud security architecture [1]. Some organizations are using the intrusion detection system (IDS) for both Host Based and Network Based in the Cloud computing [2].

The security protocols implemented to identify intrusions can be broadly summarized into the following:

Intrusion detection systems (IDS) which are hardware and/or software mechanisms that detect and log inappropriate, incorrect, or anomalous activities and report these for further investigations [3].

Intrusion Prevention Systems (IPS), which contain IDS functionality but more sophisticated systems that are capable of taking immediate action in order to prevent or reduce the malicious behavior [4]. Thus, this work utilizes both systems: (IDS) and (IPS) and refers to it as Intrusion Detection and Prevention System (IDPS). Furthermore, many works have been done in using one of the (IDS) techniques; either Anomaly Detection (AD) or Signature based Detection or hybrid of both. The ADS (Anomaly Detection System) can be used to detect unknown attacks in the networks which come from rogue nodes. In fact, such system is designed for the offline analysis due to their expensive processing and memory storage. On the other hand, the SD is used in this system to detect and identify manually the attack signature which is known as attacks in the real time traffic [5]. Therefore, both methods are essential in detecting the intrusions. So, we propose an integrated scheme which makes use of both methods to detect the attacks as soon as possible and prevent the attackers from generating the malicious activities inside the Cloud.

The rest of the paper is organized as follows: Section 2 presents a background of Cloud computing. Section 3 introduces Intrusion Detection system (IDS) with its types. We describe Intrusion Prevention System (IPS) and its difference from IDS in Section 4. In section 5, we present the IDS methods that can be used in the Cloud and where they should be deployed within the Cloud environment. Section 6 reviews the related works and distinguishes the proposed system from the previous solutions. Our proposed integrated scheme is described in section 7. A discussion of our proposed system has been reported in section 8. Finally, we conclude this paper in section 9 alongside noting down possible future research works.

## 2. Cloud Computing

Cloud computing refers to the provision of computational resources on demand via a computer network (Figure 1). Users or clients can submit a task, such as word processing, to the service provider, such as Google, without actually possessing the required software or hardware. The consumer's computer may contain very little software or data (perhaps a minimal operating system and web browser only), serving as little more than a display terminal connected to the Internet. Since the Cloud is the underlying delivery mechanism, Cloud based applications and services may support any type of software application or service in use today [6]. The essential characteristics of Cloud Computing include [7]:

1. On-demand self-service that enables users to consume computing capabilities (e.g., applications, server time, network storage) as and when required.
2. Resource pooling that allows combining computing resources (e.g., hardware, software, processing, network bandwidth) to serve multiple consumers - such resources being dynamically assigned.
3. Rapid elasticity and scalability that allow functionalities and resources to be rapidly and automatically provisioned and scaled.
4. Measured provision to optimize resource allocation and to provide a metering capability to determine usage for billing purposes Extension to existing hardware and application resources, thus, reducing the cost of additional resource provisioning.

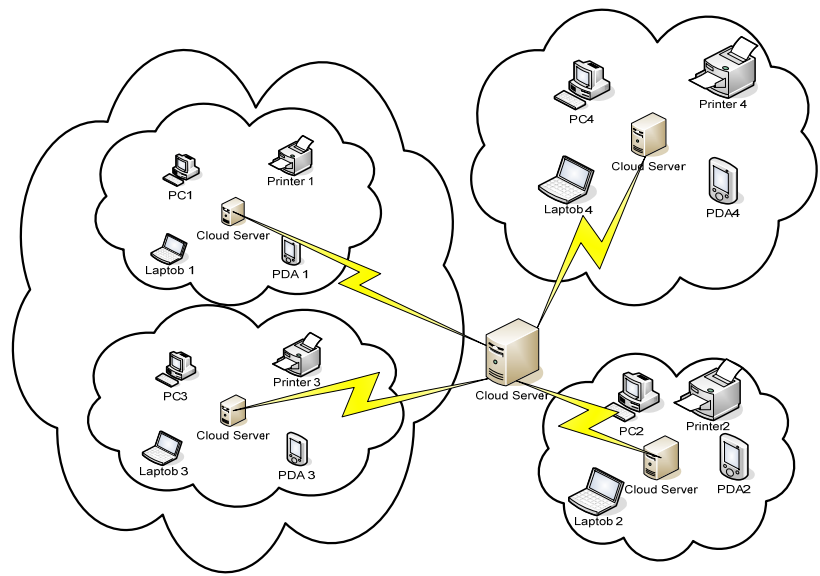

**Figure 1: Complex Cloud Computing Architecture.**

Cloud computing comprises of two different services components for the users namely as software and hardware over the Internet .However , there are various Cloud service delivery models that are developed, which can be divided into three layers [8] depending on the type of resources provided by the Cloud, distinct layers can be defined (see Figure 2). The bottom-most layer provides basic infrastructure components such as CPUs, memory, and storage, and is henceforth often denoted as Infrastructure as a Service (**IaaS**). Amazon's Elastic Compute Cloud (EC2) is a prominent example for an IaaS offer. On top of IaaS, more platform-oriented services allow the usage of hosting environments tailored to a specific need. Google App Engine is an example for a Web platform as a service (**PaaS**) which enables to deploy and dynamically scale Python and Java based Web applications. Finally, the top-most layer provides the users with ready to use applications also known as Software as a Service (**SaaS**) [8] [9].

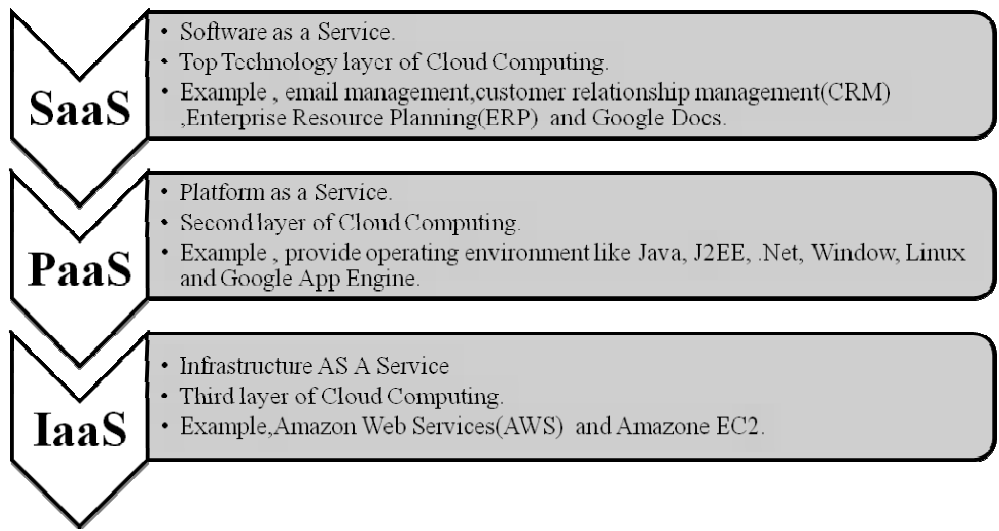

**Figure 2: Layers in Cloud computing.**

In addition, it is possible to observe the significant interaction between the services model in the Cloud computing which are Software as a service (SaaS), Platform as a service (PaaS) and Infrastructure as a Service (IaaS) as shown in Figure 3. Each one of these models provides unique service to the users in the Cloud computing environment.

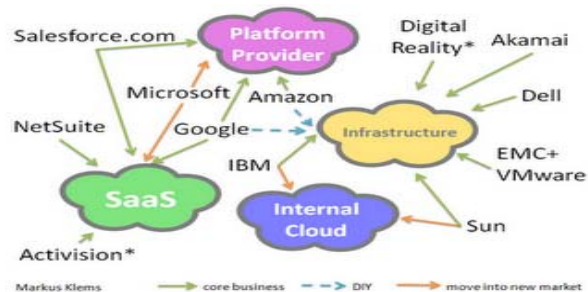

**Figure 3: Conceptual View of Services Model in the Cloud computing (Source:**

**http://ads4good.org/blog/2009/12/04/ads4good-in-the-cloud/**)

## 3. Intrusion Detection System

Intrusion detection systems (IDS) are an essential component of defensive measures protecting computer systems and network against harm abuse [1]. It becomes crucial part in the Cloud computing environment. The main aim of IDS is to detect computer attacks and provide the proper response [10]. An IDS is defined as the technique that is used to detect and respond to intrusion activities from malicious host or network [2]. There are mainly two categories of IDSs, which are listed in Table 1.

**Table 1: Types of IDS**

| Host level (HIDS) | Network level (NIDS) |
|---|---|
| Host Based | Networked and Network Node |
| Hybrid | Hybrid |
| File Integrity | Honeypots |

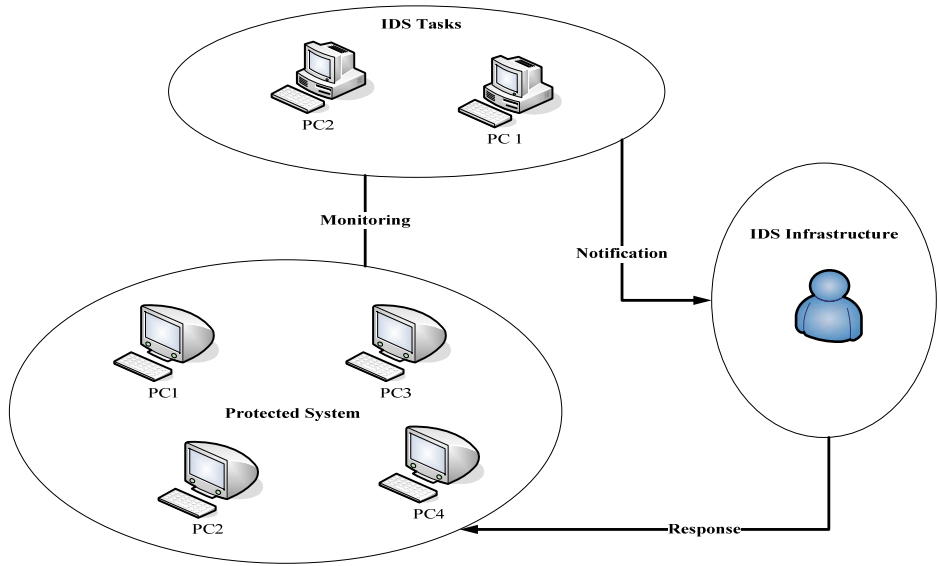

**Figure 4: Intrusion Detection System (IDS) Infrastructure.**

In addition, the IDS can be defined as a defense system, which detects hostile activities in a network. The key is to detect and possibly prevent activities that may compromise system security, or some hacking attempt in progress including reconnaissance/data collection phases that involve for example, port scans. One key feature of intrusion detection systems is their ability to provide a view of unusual activity and to issue alerts notifying administrators and/or blocking a suspected connection. Intrusion detection is defined as the process of identifying and responding to malicious activity targeted at computing and networking resources. In addition, IDS tools are capable of distinguishing between insider attacks originating from inside the organization (coming from own employees or customers) and external ones (attacks and the threat posed by hackers) [9].

Once an intrusion has been detected, IDS issues alerts notifying administrators of this fact. The next step is undertaken either by the administrators or the IDS itself, by taking advantage of additional countermeasures (specific block functions to terminate sessions, backup systems, routing connections to a system trap, legal infrastructure etc.) – following the organization's security policy (Figure 4). An IDS is an element of the security policy. Among various IDS tasks, intruder identification is one of the fundamental ones. It can be useful in the forensic research of incidents and installing appropriate patches to enable the detection of future attack attempts targeted on specific persons or resources.

### 3.1 Host Based Intrusion Detection System (HIDS)

This type of IDS involves software or agent components, which is run on the server, router, switch or network appliance. However, the agent versions must report to a console or can be run together on the same host as depicted in Figure 5. Basically, HIDS provides poor real-time response and cannot effectively defend against one-time catastrophic events. In fact, HIDSs are much better in detecting and responding to long term attacks such as data thieving [11].

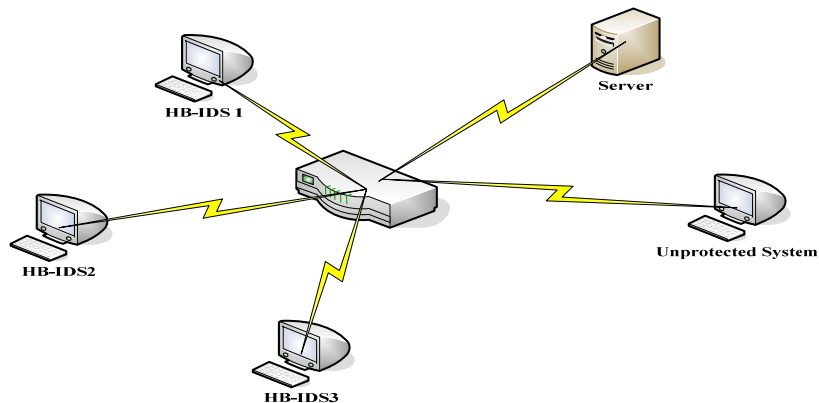

**Figure 5: Host-Based IDS.**

### 3.2 Network Based Intrusion Detection system(NIDS)

This type of IDS captures network traffic packets such as TCP, UDP and IPX/SPX) and analyzes the content against a set of RULES or SIGNATURES to determine if a POSSIBLE event took place. False positives are common when an IDS system is not configured or "tuned" to the environment traffic it is trying to analyze [12]. Figure 6 shows the network based Intrusion Detection System architecture.

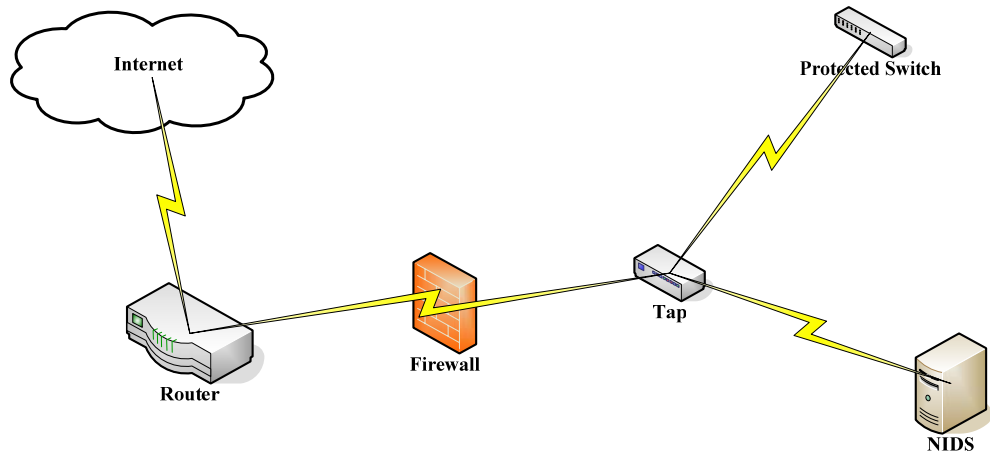

**Figure 6: Network-Based IDS.**

Table 2 summarizes the differences between the Host based Intrusion Detection system (HIDS) and Network Based Intrusion Detection System. However, the host-based and network-based systems are both required in the Cloud computing environment because they offer significantly different benefits. For an IDS, we need to use detection, deterrence, response, damage assessment, attack anticipation, and prosecution [11] [13].

**Table 2: The differences between HIDS and NIDS**

| NIDS | HIDS |
| --- | --- |
| Broad in scope | Narrow in scope |
| Easier setup and configure | More complex setup and configuration |
| Better for detecting attacks from the Outside. | Better for detecting attacks from the inside. |
| Less expensive to implement | More expensive to implement |
| Detection is based on what can be recorded on the entire network | Detection is based on what any single host can record |
| Examines packet headers | Does not see packet headers |
| OS-independent | OS-specific |
| Detects network attacks as payload is analyzed | Detects local attacks before they hit the network |
| Detects unsuccessful attack attempts | Verifies success or failure of attacks |

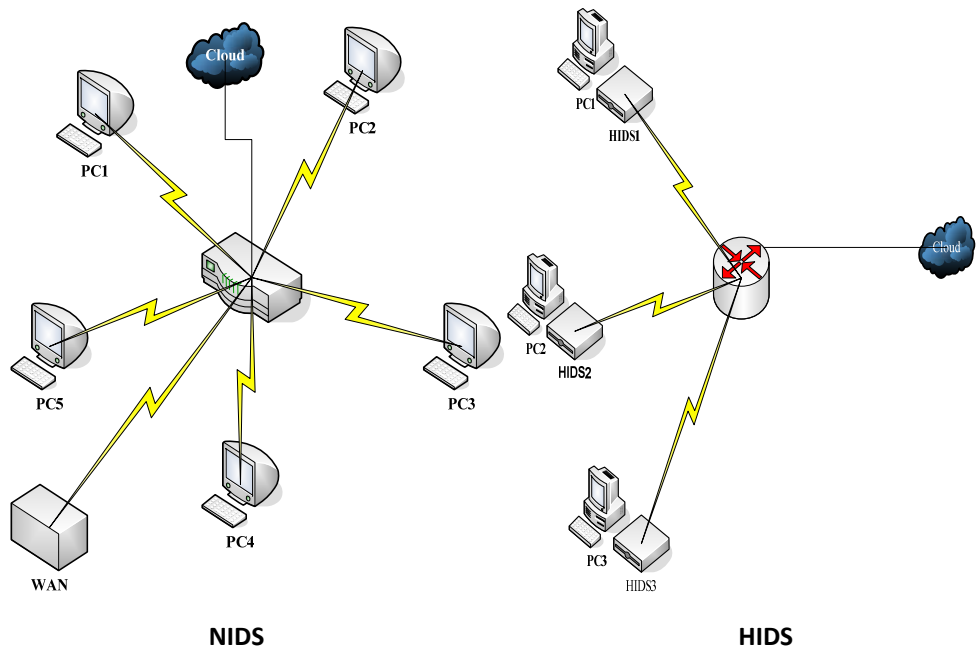

**NIDS**           **HIDS**

**Figure 7: NIDS/HIDS differences**

As has been discussed earlier about IDS and its advantage, Figure 8 shows the framework of the IDS activities. However, the main task of IDS is defending a computer system by detecting an attack and possibly repealing it. Detecting hostile attacks depends on the number and type of appropriate actions (Figure 8). Intrusion prevention requires a well-selected combination of "baiting and trapping" aimed at both investigations of threats. Diverting the intruder's attention from protected resources is another task. Both the real system and a possible trap system are constantly monitored. Data generated by intrusion detection systems is carefully examined (this is the main task of each IDS) for detection of possible attacks (intrusions). In the Table 3, we summarize the functionalities of IDS [14].

**Table 3: Functionalities of IDS [14]**

| | |
|---|---|
| 1 | Monitoring and analyzing both user and system activities. |
| 2 | Analyzing system configurations and vulnerabilities. |
| 3 | Assessing system and file integrity. |
| 4 | Ability to recognize patterns typical of attacks. |
| 5 | Analysis of abnormal activity patterns. |
| 6 | Tracking user policy violations. |

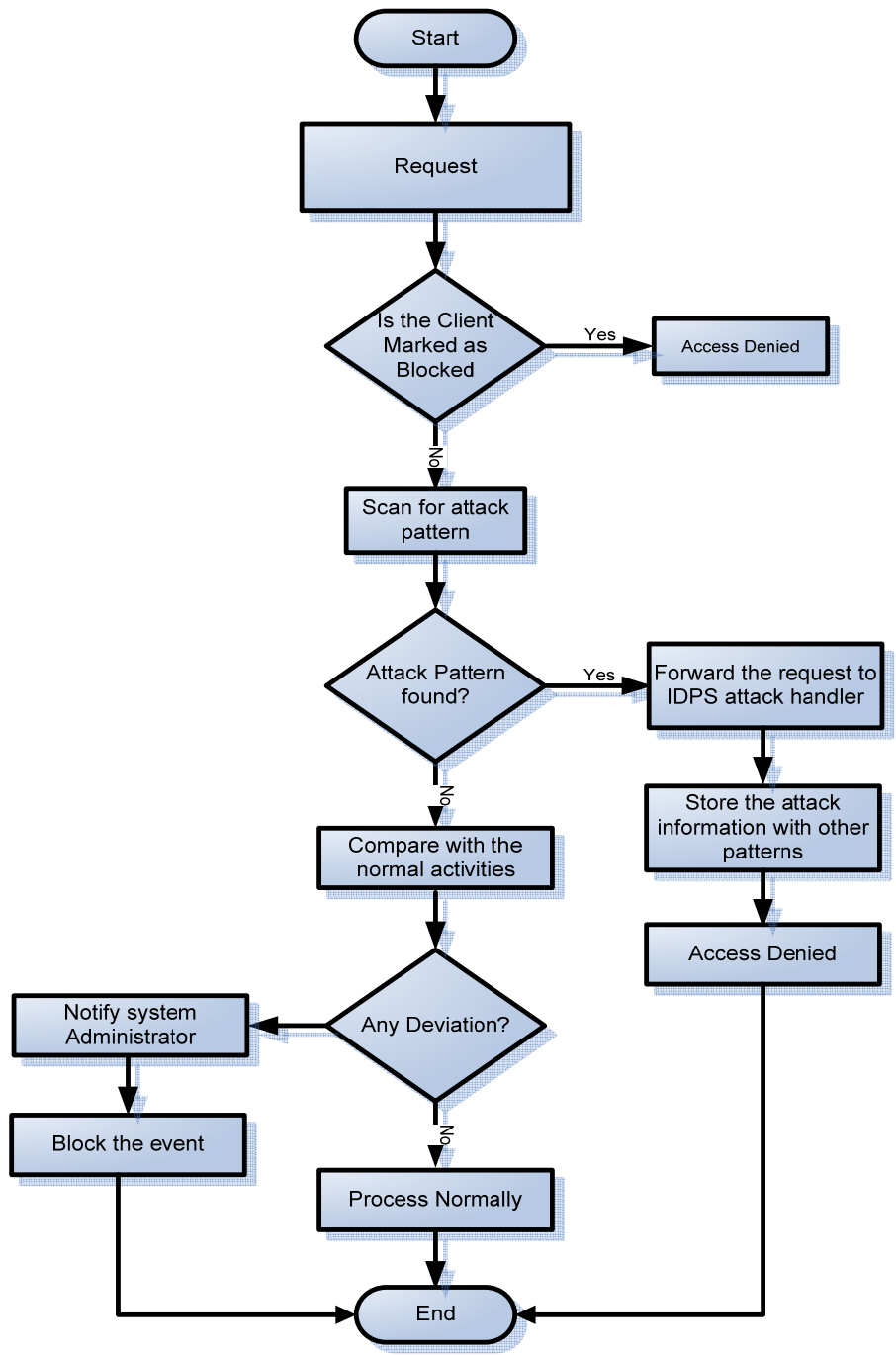

**Figure 8: IDPS Activities Framework**

## 4 Intrusion Prevention System in Cloud Computing

There are only a few works that talked about IPS in Cloud Computing. Generally an IPS sits inline on the network and monitors it, and when an event occurs, it takes action, based on prescribed rules. This is unlike IDS, which does not sit inline and is passive. Because IPSs take detection a step further, some see them as next generation IDS systems. Others, however, think in broader terms and consider the IPSs as yet another tool in the security infrastructure that could help prevent intrusions. IPS has been developed out of IDS but, the two are really different security products that have different functionalities and strengths as shown in Table 4 [15].

**Table 4: IDS vs IPS**

| IDS | IPS |
|---|---|
| Installed on network segments (NIDS) and on host (HIDS) | Installed on network segments (NIDS) and on host (HIDS) |
| Sits on network passively | Sits inline (not passive) |
| Cannot parse encrypted traffic | Better at protecting applications |
| Central management control | Central management control |
| Better at detecting hacking attacks | Ideal for blocking web defacement |
| Alerting product (reactive) | Blocking product (proactive) |

## 5. Intrusion Detection System in Cloud Computing

As mentioned before, Intrusion detection is the process of monitoring the events occurring in a computer system or network and analyzing them for signs of intrusions, defined as attempts to compromise the confidentiality, integrity, availability, or to bypass the security mechanisms of a computer or network. Intrusions are caused by attackers accessing the systems from the Internet or by authorized users of the systems who attempt to gain additional privileges for which they are not authorized or by authorized users who misuse the privileges given to them. Intrusion Detection Systems (IDSs) are software or hardware products that automate this monitoring and analysis process [16].

The Intrusion Detection Service (IDS) service increases a Cloud's security level by providing two methods of intrusion detection. First method is behavior-based method which dictates how to compare recent user actions to the usual behavior. The second approach is knowledge-based method that detects known trails left by attacks or certain sequences of actions from a user who might represent an attack. The audited data is sent to the IDS service core, which analyzes the behavior using artificial intelligence to detect deviations. This has two subsystems namely analyzer system and alert system.

In order to detect the intruders the following techniques should be implemented in either HIDS or NIDS [16] [2].

**5.1 Anomaly Detection (AD).**

Basically, Anomaly Detection was introduced in the late of 1980's with Intrusion detection expert system (IDES) [17]. Anomaly detectors identify abnormal unusual behavior (anomalies) on a host or network. They function on the assumption that attacks are different from "normal" (legitimate) activity and can therefore be detected by systems that identify these differences. Anomaly detectors construct profiles representing normal behavior of users, hosts, or network connections. These profiles are constructed from historical data collected over a period of normal operation. The detectors then collect event data and use a variety of measures to determine when monitored activity deviates from the norm. There are many measures and techniques that are used in anomaly detection including; Threshold detection, Statistical measures, Rule-based measures, other measures, including neural networks, genetic algorithms, and immune system models [16].

**5.2 Signature Detection (SD).**

Misuse detectors analyze system activity, looking for events or sets of events that match a predefined pattern of events that describe a known attack. As the patterns corresponding to known attacks are called signatures, misuse detection is sometimes called "signature-based detection". The most common form of misuse detection used in commercial products specifies each pattern of events corresponding to an attack as a separate signature. However, there are more sophisticated approaches to doing misuse detection (called "state-based" analysis techniques) that can leverage a single signature to detect groups of attacks [16]. Misuse detection techniques, in general, are not effective against the latest attacks that have no matched rules or pattern yet.

In this work, we will focus on applying IDS on IaaS which is the most flexible model for ID deployment. So, we need to identify the locations that should be considered when thinking about ID in the IaaS Cloud. There are four primary "spots" [18]:

- In the virtual machine (VM) itself: Deploying ID in the VM allows monitoring the activity of the system, and detecting and alerting on issues that may arise.
- In the hypervisor or host system: Deploying ID in the hypervisor allows to not only monitor the hypervisor but anything traveling between the VMs on that hypervisor. It is a more centralized location for ID, but there may be issues in keeping up with performance or dropping some information if the amount of data is too large.
- In the virtual network: Deploying ID to monitor the virtual network (i.e., the network established within the host itself) allows monitoring the network traffic between the VMs on the host, as well as the traffic between the VMs and the host. This "network" traffic never hits the traditional network.

- In the traditional network: Deploying ID here allows to monitor, detect, and alert on traffic that passes over the traditional network infrastructure

## 6. Relevant Works and Limitations

Many efforts have been taken in the area of Cloud computing and intrusion detection system but still there are more attacks that have not been detected. In [15], the researchers worked in this field to overcome the current security threats in the Cloud computing through implementing IDS in Cloud environment which is responsible of monitoring the utilization of resources for the virtual machine using data acquired from virtual machine monitors. More specifically, all monitoring operations are done outside the virtual machines so the attacker cannot modify the system in the case of tenant's instance is breached. However, there are many types of intrusions that this method cannot detect such as accessing the account of authorized users without any permission. In addition, if abnormal activity occurs, it will be detected as intrusion even if it is authorized activity. Therefore, all these gaps should be taken into account during implementing IDS within Cloud environment. In [19] the authors focused on number of critical issues related to security and privacy in Cloud computing environment from different perspectives such as data storage security, user identity in Cloud computing, and secure virtualization, etc. Moreover, they presented most of the attacks and threats against the Cloud with explanation of the most recent solutions to such attacks with their limitations to be solved. While, in [20] the authors concentrated on one problem regarding ensuring the integrity and correctness of user's data in the Cloud through proposing an efficient and resilient scheme against malicious data modification attack, and even server colluding attacks. On the other hand, many other researchers as in [21] are interested in distributing the IDS among the nodes of the grid within Cloud computing environment in order to monitor each node and alert the other nodes when an attack occurs. They proposed Grid and Cloud Computing Intrusion Detection System (GCCIDS) which is designed to cover the attacks that network- and host-based systems cannot detect. Their proposed method used the integration of knowledge and behavior analysis to detect specific intrusions. However, the proposed prototype cannot discover new types of attacks or create an attack database which must be considered during implementing IDS.

In [22], the authors proposed an efficient model that used multithreading technique for improving IDS performance within Cloud computing environment to handle large number of data packet flows. The proposed multi-threaded NIDS is based on three modules named: capture module, analysis module and reporting module. The first one is responsible of capturing data packets and sending them to analysis part which analyzes them efficiently through matching against pre-defined set of rules and distinguishes the bad packets to generate alerts. Finally, the reporting module can read alerts and immediately prepare alert report. The authors conducted simulation experiments to show the effectiveness of their proposed method and compared it with single thread which presented high performance in terms of processing and execution time. However, the problem of detecting new types of attacks still needs many works to be done. Even though, the researchers in [17] presented a hybrid method of integration between AD method and SD method which solved many of the previous problems, still there is a problem in stopping the attack rather than just detecting it. Obviously, the previous works focused in applying only one approach or techniques of IDS in the Cloud computing such as applying only AD or SD mechanism or even using a hybrid of both while, in this work we proposed an integrated scheme of these two techniques AD and SD in addition to the combination of two systems ID and IP which will be elaborated in details in the next section. The purpose of proposing such method is to get higher security level and to solve some gaps within the previous works as mentioned earlier.

## 7. Proposed Framework

There are several ways for the attackers to attack the target system and then taking advantage of the known vulnerabilities of computer systems. In fact, such attack leads to loss and disclosure of sensitive information and data stored in the computer. However, the IDS usually is placed in the layer which is after the firewall, what has been termed as defense in-depth strategy. In this paper, we propose a new way of protecting data and resources in the Cloud computing environment. It is based on the rational implementation of intrusion detection system (IDS) over the Cloud computing infrastructure. We focused on one layer of the Cloud computing which is known as Infrastructure as a service (Iaas). Moreover, we propose to deploy Intrusion detection and prevention system (IDPS) which is an integrated model that consists of two techniques (AD) and (SD). These two techniques will work cooperatively to perform an in-depth analysis on resources located on the Cloud to detect the intrusions and anomalies that may pose threat to the Cloud environment. These two types of attacks are different kinds of abnormal traffic events in an open network environment, whereas the intrusion takes place when an unauthorized access of a host computer system is attempted while an anomaly can be observed at the network connection level. Therefore, if any of these attacks has been detected by the proposed integrated scheme then it will compare it with the known threats (signatures) and produce an alarm in the case of matching according to Signature Based Detection technique. On the other hand, if it is not matched to any of the existing patterns, then the proposed model will detect it as abnormal behavior according to Anomaly based Detection Method and also produce an alarm and save that event as a new threat within the other signatures. In addition, the proposed system is provided also with prevention capabilities rather than just detection so it can further stop the attack itself as noted in the following:

- Terminate the user session that is being used for the attack
- Block access to the target (or possibly other likely targets) from the offending user account, IP address, or other attacker attribute
- Block all access to the targeted host, service, application, or other resource.

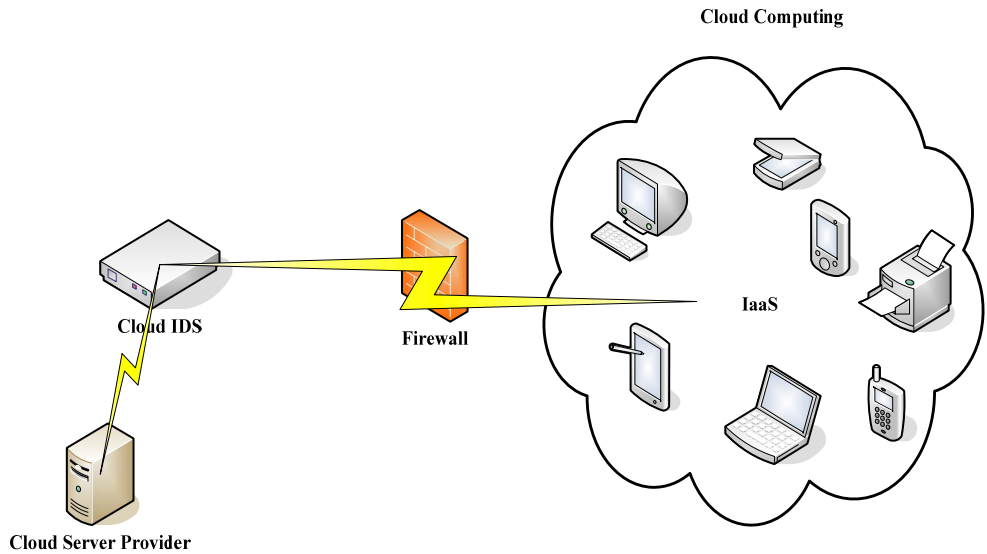

**Figure 9: The proposed Cloud IDS.**

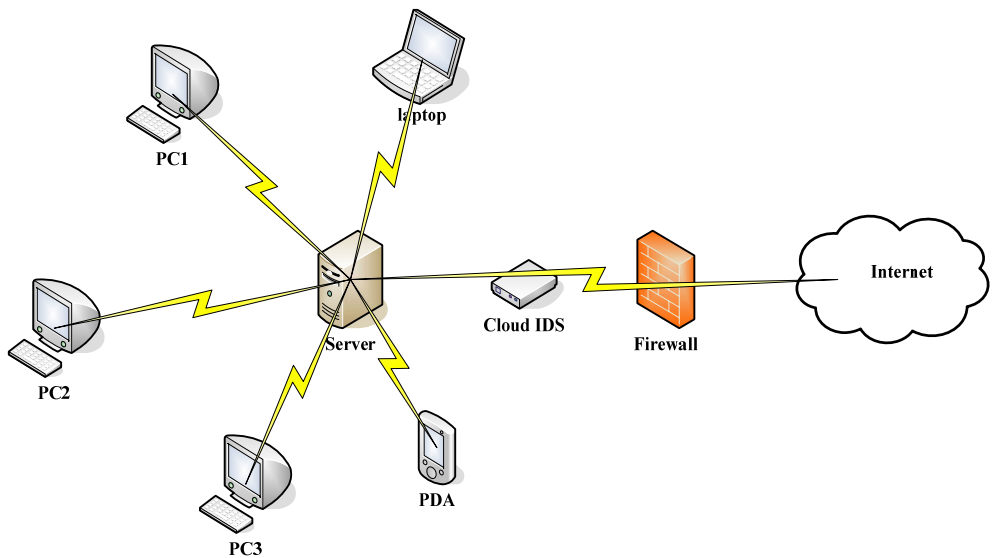

**Figure 10: Conceptual view of the Cloud IDS location.**

The integrated model uses signature matching with normal traffic profiling to enhance attack detection. Furthermore, we propose to deploy our IDS in the virtual machine itself as well as the virtual network in order to monitor the activities of the system in addition of monitoring the packet traffic in the network to filter the malicious packets coming from untrusted sources (see Figure 9). The fact is that in the Cloud computing most of the resources will be stored and accessed on the remote servers. However, the consumers do not have to worry about the maintenance and the upgrading of the software and hardware. But, the issue is when there is a flow of the packets from one source to destination; the security in terms of data integrity will not be accurate as we have the Cloud IDS placed in specific location in the NIDS. Figure10 demonstrates the close view of our proposed method to protect the data and resources in the Cloud.

## 8. Discussion

Basically, the IDS have been implemented in organizations to collect and analyze various types of attacks within a host system or a network. In addition, to identify and detect possible threats violations, which involve both intrusions, which are the attacks from outside the organizations and misuses that are known as the attacks within the organizations. In this paper, we proposed the integrated model which involves a combination of the two systems Intrusion Detection (ID) and Intrusion Prevention (IP) adding to those getting benefits from integrating of the two known techniques: anomaly Detection (AD) and Signature Detection (SD), which is totally different from most of the recent works that focused only on using one system, either detection or prevention and also using either Anomaly detection or Signature based detection. Some works even used a hybrid method which is a combination of both such as the work presented in [17] where the researchers used the merger between both methods (AD) and (SNORT) which is ID based on Signature but even then, their method was not provided with prevention capabilities. On the other hand, in our case, we proposed to use our approach IDPS, which not only can detect the attack but also can further stop it using the capabilities of prevention system which has not been utilized in the previous works. Therefore, the proposed system can outperform the hybrid system of [17] in terms of preventing the attack from conducting any bad action through blocking the event and saving that threat with the other signatures in order to be observed by Signature Based Intrusion Detection for next time so that it can be detected earlier.

Finally, deploying such integrated model in the Cloud environment will reduce the probability of risks than the normal system or even than other systems which are just provided with Intrusion Detection methods.

## 9. Conclusions and Future Works

Cloud computing has motivated the introduction of a new service to the Information Technology (IT) discipline. The use of Cloud computing will reduce the infrastructure maintenance cost, scalability for data and applications, availability of data services and pay as you use features. Since the idea of Cloud computing is well known as a network of networks over the World Wide Web, consequently, the probability of having various types of vulnerabilities causing attacks is high. Keeping this fact in mind, in this paper we discussed different techniques of an intrusion detection system that has been used to counter malicious attacks in Cloud computing environment. For Cloud computing, several network access rates are used and control of data & applications are needed for each service provider. Therefore, an efficient, reliable and information transparent IDS is required.

Many researchers think that using AD could provide reasonable level of security for the Cloud while, others think that using SD may provide better security. In fact, both methods are very important for deploying IDS in the Cloud and they complement each other. Therefore, we have proposed a method of combining both techniques as an integrated IDS technique to benefit from both of these techniques in detecting as much attacks as possible.

Our proposed system is provided with prevention capabilities which make it unique among other previous solutions in terms of stopping the attack rather than just detecting or reporting alarms. For future research work, we suggest to do the implementation of our proposed IDPS approach in a real Cloud computing environment to verify our envisioned outcome. Also, we plan to deploy a honeypot in the proposed architecture to ensure good performance, we wish to increase the level of security in the Cloud computing environment and decrease the threats to Cloud environments through focusing on the problem of how data are stored in the Cloud.